\documentclass[print,twocolumn]{revtex4}
\usepackage[dvips]{graphicx}
\usepackage{graphicx}
\usepackage{dcolumn}
\usepackage{bm}
\usepackage{epsfig}
\usepackage{setspace}

\begin{document}

\title{Entanglement of Two-Superconducting-Qubit System Coupled with a Fixed Capacitor}

\author{TIAN Li-Jun $^{1,2}$}\email{tianlijun@staff.shu.edu.cn}
\author{ QIN Li-Guo
{$^{1,2}$}} \email{lgqin@shu.edu.cn}
\author{ZHANG Hong-Biao $^3$}

\address{$^{1}$ Department of Physics, Shanghai University, Shanghai 200444\\
$^{2}$ Shanghai Key Lab for Astrophysics, Shanghai 200234 \\
$^{3}$ Institute of Theoretical Physics, Northeast Normal
University, Changchun 130024 }

\begin{abstract}
We study thermal entanglement in a two-superconducting-qubit system
in two cases, either identical or distinct. By calculating the
concurrence of system, we find that the entangled degree of the
system is greatly enhanced in the case of very low temperature and
Josephson energies for the identical superconducting qubits, and our
result is in a good agreement with the experimental data.

PACE: 03.65.Ud, 74.81.$-$g, 74.50.+r

Keywords: Quantum Entanglement; Superconducting qubits; Josephson
energy

\end{abstract}
 \maketitle

Entanglement has received much attention since it plays a central
role in quantum information processing and quantum
computing.\cite{mni} There are several ways to generate entanglement
through experiments.\cite{yap,ajb} However, it is still an open
question to generate very good entangled states. Current interest
focuses on generating, maintaining and controlling precisely
entanglement of systems.\cite{csy,kcs} As is well known, temperature
and magnetic field can be prepared, control and maintain
entanglement.\cite{mcs,ams} One may ask a question: are there any
other effective ways to control entanglement?

For micro-systems, entanglement between different kinds of qubits
has been studied, for example, charge,\cite{ldc,tyy} flux,\cite{jbm}
charge flux,\cite{dva} and phase qubits.\cite{jmm,rmr} Of particular
importance, the superconducting qubits\cite{jqu} take advantage of
the two characteristic of superconducting and quantum, and therefore
become the most suitable candidates for quantum computing,\cite{pld}
which has been carried out in laboratory.\cite{hmo} For
superconducting qubits, manipulation of quantum states has enabled
scientists to generate partly entangled states.\cite{aim} However,
high quantum entangled states are required in such quantum
technology. In the experimental aspect, entanglement has been
generated for coupled charge qubits\cite{yap} and coupled phase
qubits,\cite{ajb} but the maximally entangled states are merely in
theory.\cite{mdk} On the other hand, recent experiments have
observed strong couplings between two superconducting
qubits.\cite{rmr,awd,pya} As a consequence, they triggers the
theoretical research on investigating superconducting qubits. We
have carried out some research on the corresponding relations
between the theory and experiment in quantum entanglement.\cite{ljt}

In this Letter, based on experimental study of changing the
entanglement degree by adjusting the capacitance and LC
circuits,\cite{mds} thermal entanglements are studied in two
superconducting qubits, either identical or distinct. Different
evolutions of the entanglement are observed. In the case that the
superconducting qubits have the same Josephson energies, we
investigate the effect of temperature and Josephson energies on
entanglement. The result exhibits high quantum entangled states at
low temperature. In addition, our theoretical results match with the
experimental data very well, so the entangled qubits, which are made
by making use of our data, should have better entangled nature. We
hope that this would be confirmed experimentally in the future.

The present model is composed of two single cooper-pair box charge
qubits, coupled with a fixed capacitor. This model has attracted
much attention and researches.\cite{mds,epa,gsp,jlk} The
superconducting materials act as a superconductor with a suppressed
transition temperature $T_{c}$ adjusted by using different
materials, which is in mK range for practical operation in efficient
and multiplex superconducting circuits. One good example is a
superconductive material made from a superconducting Al/Ti/Au
trilayer with respective thicknesses of $300$, $200$, and $200$
{\AA}, $T_{C}=450$\,mK.\cite{mds,pme} The Hamiltonian of
two-superconducting-qubit system is given by\cite{mds}
\begin{eqnarray}
\label{lg1}
H=&&-\frac{1}{2}\{[4E_{C1}(\frac{1}{2}-n_{g1})+2E_{m}(\frac{1}{2}-n_{g2})]\sigma_{z1}\nonumber\\
&&+[4E_{C2}(\frac{1}{2}-n_{g2})+2E_{m}(\frac{1}{2}-n_{g1})]\sigma_{z2}\nonumber\\&&+E_{J1}\sigma_{x1}+E_{J2}\sigma_{x2}-2E_m\sigma_{zz}\},
\end{eqnarray}
where $E_{Cj}$ and $E_{Jj}$ are respectively the charging and
Josephson energies, and $E_{m}$ is the mutual coupling energy
between the two qubits; $\sigma_{x1}=\sigma_{x} \otimes I$,
$\sigma_{x2}=I \otimes\sigma_{x}$, $\sigma_{zz}=\sigma_{z} \otimes
\sigma_{z}$ with $\sigma_{x,z}$ being the normal Pauli matrices and
$I$ the identity matrix; $n_{gj}=C_{gj}V_{gj}/2e$ is the normalized
qubit gate charge with $C_{gj}$ and $V_{gj}$ the control gate
capacitance and voltage, respectively.

For simplicity, calculations are restricted at the degeneracy point,
where $n_{g1}=n_{g2}=0.5$, which is the condition of insensitivity
to noise.\cite{gsp} Under this condition, the model Hamiltonian
reduces to
\begin{eqnarray}
\label{lg2}
H=-\frac{1}{2}(E_{J1}\sigma_{x1}+E_{J2}\sigma_{x2}-2E_m\sigma_{zz}),
\end{eqnarray}
which is independent of charging energy. This reduced Hamiltonian is
applied to study quantum gates too,\cite{epa} The eigenvalues and
eigenvectors of Hamiltonian can be obtained,
\begin{eqnarray}
\label{lg3}
&&H|\Psi_{1}\rangle=-\frac{1}{2}\sqrt{A}|\Psi_{1}\rangle,
H|\Psi_{2}\rangle=\frac{1}{2}\sqrt{A}|\Psi_{2}\rangle,\nonumber\\
&&H|\Psi_{3}\rangle=\frac{1}{2}\sqrt{B}|\Psi_{3}\rangle,
H|\Psi_{4}\rangle=-\frac{1}{2}\sqrt{B}|\Psi_{4}\rangle,
\end{eqnarray}
where
\begin{eqnarray}\label{llg1}
&&|\Psi_{1}\rangle = \frac{1}{N_{1}}[|00\rangle - |11\rangle -
a_{1}(|01\rangle - |10\rangle)],\nonumber\\
&&|\Psi_{2}\rangle = \frac{1}{N_{2}}[|00\rangle - |11\rangle +
a_{2}(|01\rangle -
|10\rangle)],\nonumber\\
&&|\Psi_{3}\rangle = \frac{1}{N_{3}}[(|00\rangle + |11\rangle)+
a_{3} (|01\rangle+|10\rangle)],\nonumber\\
&&|\Psi_{4}\rangle = \frac{1}{N_{4}}[(|00\rangle + |11\rangle)-
a_{4}(|01\rangle+|10\rangle)],
\end{eqnarray}
$A=(E_{J1}-E_{J2})^2+4E^{2}_{m}$ and
$B=(E_{J1}+E_{J2})^2+4E^{2}_{m}$.
 Here
$a_{1}=(\sqrt{A}+2E_{m})/(E_{J1}-E_{J2})$,
$a_{2}=(\sqrt{A}-2E_{m})/(E_{J1}-E_{J2})$,
$a_{3}=(\sqrt{B}+2E_{m})/(E_{J1}+E_{J2})$, and
$a_{4}=(\sqrt{B}-2E_{m})/(E_{J1}+E_{J2})$. $N_{i}$ is the
normalization coefficient of $|\Psi_{i}\rangle$ $(i=1, 2, 3, 4)$.
For $E_{J1}E_{J2}>0$, the ground state is $|\Psi_{4}\rangle$, and
$|\Psi_{1}\rangle$ for $E_{J1}E_{J2}<0$. An important observation is
that for the attractive case of $E_{J1}E_{J2}=0$, the degeneracy
states in the ground state appear.

In order to measure entanglement, concurrence has been proposed, and
is defined as\cite{sh,wk}
\begin{eqnarray}
\label{lg7} C = max\{\lambda_{1} - \lambda_{2} -\lambda_{3}
-\lambda_{4}, 0\},
\end{eqnarray}
where the parameters $\lambda_{i}$ in decreasing order are the
square roots of the eigenvalues of the operator
\begin{eqnarray}
\label{lg8}
\varsigma=\rho(\sigma^{y}_{1}\otimes\sigma^{y}_{2})\rho^{*}(\sigma^{y}_{1}\otimes\sigma^{y}_{2}),
\end{eqnarray}
where $\sigma^{y}_{1,2}$ are the Pauli spin matrix of the two qubits
and $\rho=(1/Z)\exp(-H/kT)$ is the density operator of the system at
the thermal equilibrium, where $Z={\rm Tr}[\exp(-H/kT)]$ is the
partition function. The concurrence $C$ ranges from $0$ for a
separable state to $1$ for a maximally entangled state. Following
the same method in the standard basis,
the density matrix of the system is \nonumber\\
$ \label{lg9}\rho(T)=\frac{1}{Z} \left( \begin {array}{cccc}
m_{{1}}-m_{{2}}&m_{{3}}-m_{{4}}&m_{{3}}+m
_{{4}}&m_{{5}}+m_{{6}}\\\noalign{\medskip}m_{{3}}-m_{{4}}&m_{{1}}+m_{{
2}}&m_{{5}}-m_{{6}}&m_{{3}}+m_{{4}}\\\noalign{\medskip}m_{{3}}+m_{{4}}
&m_{{5}}-m_{{6}}&m_{{1}}+m_{{2}}&m_{{3}}-m_{{4}}\\\noalign{\medskip}m_
{{5}}+m_{{6}}&m_{{3}}+m_{{4}}&m_{{3}}-m_{{4}}&m_{{1}}-m_{{2}}
\end {array} \right)$,

\noindent where $m_{1}=\frac{1}{2}(\cosh(x_{A})+\cosh(x_B))$,
$m_{2}=
E_{m}(\frac{\sinh(x_A)}{\sqrt{A}}+\frac{\sinh(x_B)}{\sqrt{B}})$,
$m_{3}=\frac{E_{J1}+E_{J2}}{2\sqrt{B}}\sinh(x_B)$,
$m_{4}=\frac{E_{J1}-E_{J2}}{2\sqrt{A}}\sinh(x_A)$,
$m_{5}=\frac{1}{2}(\cosh(x_B)-\cosh(x_A))$,
$m_{6}=E_{m}(\frac{\sinh(x_A)}{\sqrt{A}}-\frac{\sinh(x_B)}{\sqrt{B}})$
and $Z= 4m_{1}$, with $x_A=\frac{\sqrt{A}}{2kT}$ and
$x_B=\frac{\sqrt{B}}{2kT}$. The concurrence can be easily calculated
by Eqs.\,({\ref{lg7}) and (\ref{lg8}).

For identical superconducting qubits, $E_{J1}=E_{J2}=E_{J}$ and
$E_{C1}=E_{C2}$, the model Hamiltonian can be rewritten as
\begin{eqnarray}
\label{lg4}
H=-\frac{1}{2}(E_{J}\sigma_{x1}+E_{J}\sigma_{x2}-2E_m\sigma_{zz}),
\end{eqnarray}
where $E_{J}$ and $E_m$ can be adjusted by the experimental
multiplexed capacitance in the circuits. Similar model was argued
elsewhere for the choices of $E_J$ as a magnetic field.\cite{gdv}

The eigenvalues and eigenvectors of Hamiltonian Eq. (\ref{lg4}) read
\begin{eqnarray}
\label{lg5} &&H|\psi_{1}\rangle=-E_{m}|\psi_{1}\rangle,
H|\psi_{2}\rangle=E_{m}|\psi_{2}\rangle,\nonumber\\
&&H|\psi_{3}\rangle=\sqrt{D}|\psi_{3}\rangle,
H|\psi_{4}\rangle=-\sqrt{D}|\psi_{4}\rangle
\end{eqnarray}
with $D=E^{2}_{m}+E^{2}_{J}$; $|\psi_{1}\rangle =
\frac{1}{\sqrt{2}}(|01\rangle - |10\rangle)$, $|\psi_{2}\rangle =
\frac{1}{\sqrt{2}}(|00\rangle - |11\rangle)$, $|\psi_{3}\rangle =
\frac{1}{N_{+}}[(|00\rangle +
|11\rangle)+\xi_{-}(|01\rangle+|10\rangle)]$, and $|\psi_{4}\rangle
= \frac{1}{N_{-}}[(|00\rangle +
|11\rangle)+\xi_{+}(|01\rangle+|10\rangle)]$, with
$\xi_{\pm}=\frac{E_{m}\pm\sqrt{D}}{E_{J}}$ and $N_{\pm}$ are the
normalization coefficients. Here $|\psi_{1}\rangle$ and
$|\psi_{2}\rangle$ are two of four Bell states, which are the
maximally entangled states.

The density matrix can be obtained by the above same way in this
case. Without lack of generality, the mutual coupling energy $E_{m}$
is regarded as the energy unit and $k=1$. Thus we can consider
$E_{m}/k=1$, whose unit is mK. For convenience, we only write its
value as the same as $E_{J}$. By making use of Eqs. ({\ref{lg7}})
and ({\ref{lg8}}), the concurrence can be calculated for the
identical qubits. Especially, for $E_J=0$ the Hamiltonian
(\ref{lg4}) only has the last item whose eigenvectors are the
separable states, so that no thermal entanglement is present,
namely, $C=0$.

\begin{figure}[th]
\centerline{\psfig{file=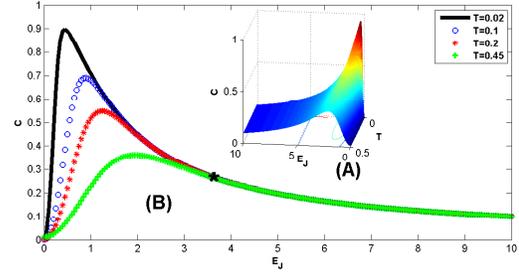,width=8cm}} \caption{(Color online)
Two-dimensional plots of the concurrence vs Josephson energy $E_{J}$
for different temperatures. Asterisks: the experimental data. Inset:
three-dimensional plot of the concurrence as a function of $E_{J}$
and $T$.  }
\end{figure}

In the inset of Fig. 1, we show the concurrence as functions of
Josephson energies and temperature, displaying nonmonotonic behavior
for smaller $E_{J}$ and lower temperature. However, in the limit
$E_{J}\rightarrow 0$, $E_m\approx\sqrt{D}$, the degeneracy states
are present: $\psi_1$ and $\psi_4$; $\psi_2$ and $\psi_3$, namely,
only two energy levels are populated. Thus there is the energy level
crossing the point $E_J=0$, namely, the ground state is the
degenerate state of $\psi_4$ and $\psi_1$. With an infinitesimal
increase of $E_{J}$, the concurrence will increase sharply to a top
in accordance with Ref \cite{gdv}. At the zero temperature, the
entanglement primarily depends on $|\psi_{4}\rangle$, i.e. on the
ground state, which plays a major role. As $T$ increase, the peaks
fall, because the ground state will mix with excited states in
thermodynamic equilibrium and mixing states combine the concurrence
of the system. To illustrate this feature, Figs. 1 and 2 are plotted
to show the behavior of $C$ vs $E_{J}$ and $C$ vs $T$, respectively.

Figure 1 clearly shows that no entanglement is present for $E_J=0$.
As $E_J$ increases, the entanglement first reaches sharply to the
maximum, then decays rapidly, finally reduces slowly and
asymptotically to a stable value. Moreover, lower temperature and
lower Josephson energy will cause the entanglement richer.
Considering the data of the sample $2$ for the identical qubits in
Ref.\,\cite{mds} and $E_{m}$ as the energy unit, we obtain
$E_{J1}/k=E_{J2}/k=3.625$. For $T=20$\,mK and $E_{J}=3.625$, the
theoretical prediction is $C=0.26593$, which shows the excellent
agreement with $C=0.27$ observed experimentally.\cite{mds}

\begin{figure}[th]
\centerline{\psfig{file=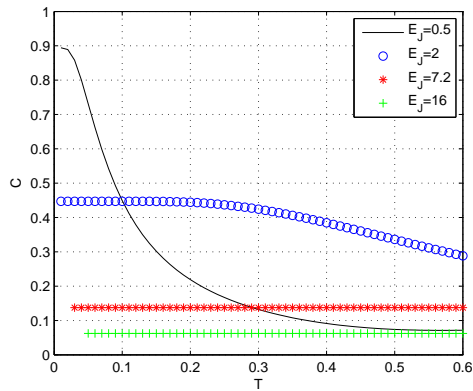,width=7cm}} \caption{(Color online)
Two-dimensional plots of the concurrence as a function of the
temperature $T$ for the four different Josephson energy.}
\end{figure}

In Fig. 2, $C$ vs $T$ for different Josephson energy is presented.
For lower Josephson energy, for example $E_{J}= 0.5$, the
concurrence will vary dramatically, but is not so apparent for
bigger Josephson energy. When the temperature $T=0$, only the ground
state $|\psi_{4}\rangle$ exists, and $C \approx 0.9$. With increase
of $E_{J}$, $C_{|\psi_{4}\rangle}$ will decrease, so the
intersections of the curve and $C$-axis decline. For a fixed smaller
$E_J$, as the temperature rising, the ground state and three excited
states mix, $C$ will decrease sharply. On the contrary, for the
larger Josephson energy, the change behavior of $C$ becomes very
slow and finally $C$ tends stably at $T\leq T_C$. Thus, the
concurrence is very susceptible to small Josephson energy at lower
temperature (see Fig. 2).

\begin{figure}[th]
\centerline{\psfig{file=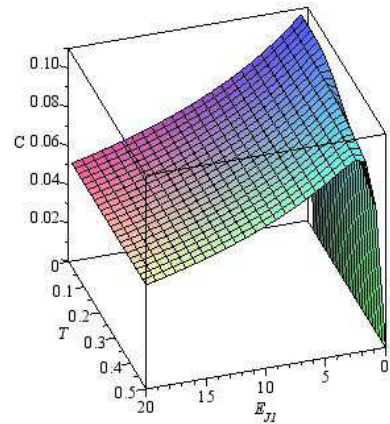,width=7cm}} \vspace*{8pt}
\caption{(Color online) Three-dimensional plot of the concurrence as
functions of temperature $T$ and Josephson energy $E_{J1}$ with
$E_{J2}=17.2$. }
\end{figure}

For the distinct superconducting qubits, the concurrence can be
calculated through Eqs. ({\ref{lg7}}) and ({\ref{lg8}}). To
distinguish different influences of identical and distinct
superconducting qubits on the entanglement at the same temperature,
Figure 3 shows the evolution of the concurrence as functions of
$E_{J1}$ and $T$ with $E_{J2}=17.2$. Obviously $C$ is smaller than
that of the identical case at low temperature.

\begin{figure}[th]
\centerline{\psfig{file=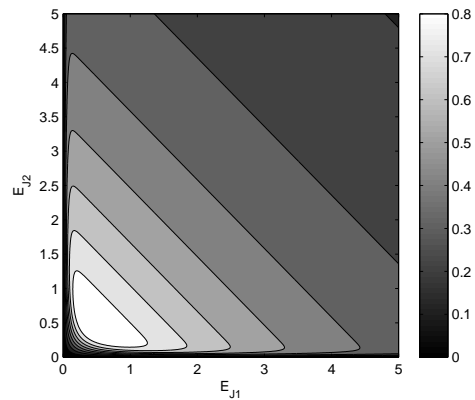,width=7cm}} \vspace*{6pt}
\caption{(Color online) Two-dimensional contour plots of the
concurrence $C$ as a function of the Josephson energies $E_{J1}$ and
$E_{J2}$ at $T=20$\,mK. }
\end{figure}

\begin{figure}[thb]
\centerline{\psfig{file=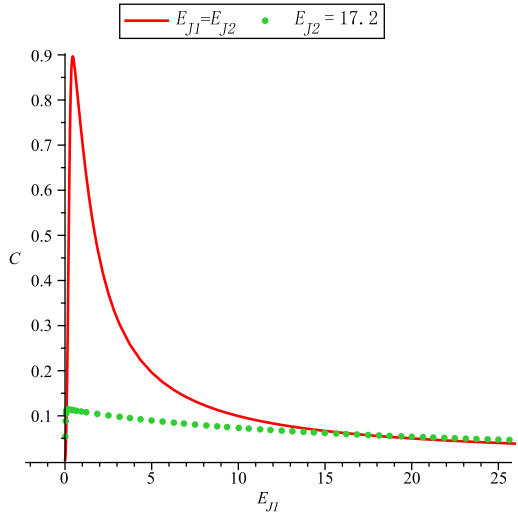,width=7cm}} \vspace*{6pt}
\caption{(Color online) Two-dimensional plots of the concurrence as
a function of Josephson energy $E_{J1}$ for $T=20$\,mK.}
\end{figure}

The contour figure of $C$ is plotted in Fig. 4. It is worth noting
that two Josephson energies are smaller and closer, the concurrence
decreases more slowly, and the peak is higher. This proves that $C$
is maximal at $E_{J1}=E_{J2}$ for the stable temperature. When the
values of $E_{J1}$ stay away from $E_{J2}$, $C$ will decay.
According to the data in Ref.\cite{mds}, by taking $E_{J1}=13.6$ and
$E_{J2}=17.2$, our theoretical result is $C=0.064$, which matches
with the experimental $C=0.06$.\cite{mds}

 To compare in more detail, evolutions of concurrences are shown in Fig. 5 for the two
cases. One can clearly find the difference between them. The
apparent difference is the maximal concurrence. The maximum value of
$C$ for $E_{J1}=E_{J2}$ is much larger than $E_{J1}\neq E_{J2}$.
That is to say, choosing the proper superconducting qubits can
enhance the entanglement at low temperature. Two Josephson energies
are smaller and closer, then the maximum value of concurrence will
be larger.

In conclusion, we have investigated the effect of Josephson energies
on the thermal entanglement in the two-superconducting-qubit model.
In the two cases, i.e., identical and distinct superconducting
qubits, we have presented the evolution of concurrence with respect
to the Josephson energy and temperature. Comparing the results of
these two cases, we conclude that the entanglement may be enhanced
under the identical superconducting qubits for the same temperature.
When the temperature and Josephson energies are lower, the entangled
degree of the system is greatly enhanced. Our theoretical prediction
is in good agreement with the experiments and provides a new way to
enhance and to control the entanglement degree of the system by
adjusting the Josephson energies, which can be realized
experimentally by changing the capacitance and LC circuits.
Utilizing the results of calculation and investigation, we may
generate better entangled and stable states, which could have wide
applications in the quantum communication and physical experiments.

This work is partly supported by the NSF of China (Grant No.
11075101), by Shanghai Leading Academic Discipline Project (Project
No. S30105), and by Shanghai Research Foundation (Grant No.
07d222020). The authors are grateful to Xin-Jian Xu and Ying Jiang
for valuable discussions.

\end{document}